\documentclass[12pt,doublespace]{article}
\usepackage{fullpage}
\usepackage{setspace}
\usepackage{hyperref}
\usepackage{graphicx}
\usepackage{cite}
\usepackage[labelfont=bf]{caption}
\makeatletter
\renewcommand\@biblabel[1]{#1.}

\makeatother

\begin{document}
\noindent
\begin{center}
\textbf{Title: Electron-Hole Asymmetric Integer and Fractional Quantum Hall Effect in Bilayer Graphene}\\
\end{center}

\noindent
\textbf{Authors: A. Kou$^{1*}$, B. E. Feldman$^{1*}$, A. J. Levin$^1$, B. I. Halperin$^{1,2}$, K. Watanabe$^3$, T. Taniguchi$^3$, A. Yacoby$^{1,2\dagger}$ }\\

\noindent
\textbf{Affiliations:}

\noindent
$^1$Department of Physics, Harvard University, Cambridge, MA 02138, USA.\\
\noindent
$^2$CIQM: Center for Integrated Quantum Materials\\
\noindent
$^3$National Institute for Materials Science, Tsukuba, Japan\\
\noindent
$^*$These authors contributed equally to this work.\\
\noindent
$^\dagger$email: yacoby@physics.harvard.edu\\

\noindent
\textbf{Abstract:} The nature of fractional quantum Hall (FQH) states is determined by the interplay between the Coulomb interaction and the symmetries of the system. The unique combination of spin, valley, and orbital degeneracies in bilayer graphene is predicted to produce novel and tunable FQH ground states. Here we present local electronic compressibility measurements of the FQH effect in the lowest Landau level of bilayer graphene. We observe incompressible FQH states at filling factors $\nu=2p+2/3$ with hints of additional states appearing at $\nu=2p+3/5$, where $p=-2,-1,0, $ and 1. This sequence of states breaks particle-hole symmetry and instead obeys a $\nu \rightarrow \nu+2$ symmetry, which highlights the importance of the orbital degeneracy for many-body states in bilayer graphene.\\

\noindent
\textbf{One Sentence Summary:} Bilayer graphene exhibits a unique sequence of electron-hole asymmetric fractional quantum Hall states because of its twofold orbital degeneracy.\\
\\
\\

\noindent
\textbf{Main Text:}\\
\indent
The charge carriers in bilayer graphene obey an electron-hole symmetric dispersion at zero magnetic field. Application of a perpendicular magnetic field $B$ breaks this dispersion into energy bands known as Landau levels (LLs). In addition to the standard spin and valley degeneracy found in monolayer graphene, the $N=0$ and $N=1$ orbital states in bilayer graphene are also degenerate and occur at zero energy \cite{McCann}. This results in a sequence of single-particle quantum Hall states at $\nu=4Me^2/h$, where $M$ is a nonzero integer \cite{GeimBilayer}.

When the disorder is sufficiently low, the eightfold degeneracy of the lowest LL is lifted by electron-electron interactions, which results in quantum Hall states at all integer filling factors \cite{befBilayer,Zhao_PRL104_066801_2010}. The nature of these broken-symmetry states has been studied extensively both experimentally and theoretically, with particular attention given to the insulating $\nu=0$ state. Multiple groups have been able to induce transitions between different spin and valley orders of the ground states using external electric and magnetic fields, which indicates the extensive tunability of many-body states in bilayer graphene \cite{Tomas, Kim_PRL107_016803_2011, Velasco, Maher}. The interplay between externally applied fields and intrinsic electron-electron interactions, both of which break the degeneracies of bilayer graphene, produces a rich phase diagram not found in any other system.

Knowledge of the ground state at integer filling factors is especially important for investigating the physics of partially filled LLs, where in exceptionally clean samples, the charge carriers condense into fractional quantum Hall (FQH) states. The above-mentioned degrees of freedom as well as the strong screening of the Coulomb interaction in bilayer graphene are expected to result in an interesting sequence of FQH states in the lowest LL \cite{Apalkov, Papic, PapicTun, Snizhko, Apalkov_PRL107_186803_2011, Papic_arXiv1307_2909}. Indeed, partial breaking of the SU(4) symmetry in monolayer graphene has already resulted in sequences of FQH states with multiple missing fractions \cite{Du, Bolotin, Dean, Lee, befSci,befPhase}.

\begin{figure}
\centering
\includegraphics[width=\textwidth]{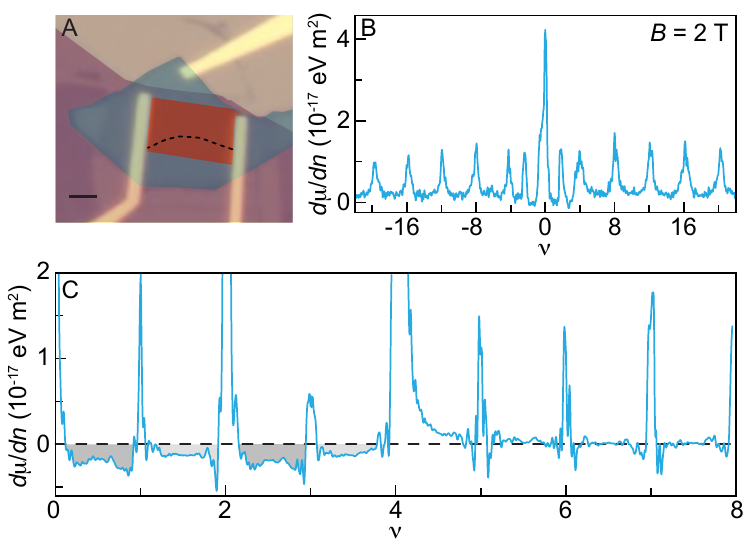}
\caption{
\label{BNGFig1}}
\end{figure}

Experimental observation of FQH states, however, has proven to be difficult in bilayer graphene. Hints of a $\nu=1/3$ state were first reported in transport by Bao and coworkers \cite{JeannieLau}. Very recently, Ki and coworkers observed robust FQH states at $\nu=-1/2$ and $\nu=-4/3$ in a current-annealed suspended bilayer sample \cite{Morpurgo}. Here, we report local compressibility measurements of a bilayer graphene device fabricated on hexagonal-Boron Nitride (h-BN), performed using a scanning single-electron transistor (SET). Our technique allows us to directly probe the thermodynamic properties of the bulk of the sample \cite{Yoo, Amir}. We measure the local chemical potential $\mu$ and the local inverse compressibility $d\mu/dn$ by changing the carrier density $n$ with a proximal graphite gate located 7.5~nm from the graphene and monitoring the resulting change in SET current. An optical image of the contacted device is shown in Fig. \ref{BNGFig1}A.

Figure \ref{BNGFig1}B shows a measurement of the inverse compressibility as a function of filling factor at $B=2$~T. Incompressible features are present at all nonzero multiples of $\nu=4$, indicating that we are measuring bilayer graphene. The full width at half maximum of the $\nu=4$ peak provides a measure of the disorder in the system and is on the order of $10^{10}~\mathrm{cm}^{-2}$, similar to that observed in suspended bilayers \cite{Jens, befBilayer}. Broken-symmetry states at $\nu=0$ and $\pm 2$ are also visible at $B=2$~T, which further indicates the cleanliness of the sample.

In Fig. \ref{BNGFig1}B, the inverse compressibility appears more negative for $|\nu|<4$ than in higher LLs, and we explore this behavior in more detail below. The average inverse compressibility between 8 and 11.5~T is plotted as a function of filling factor in Fig. \ref{BNGFig1}C.  All of the integer broken-symmetry states have fully developed at these magnetic fields, and averaging over a range of fields helps to clarify the underyling behavior of the inverse compressibility by reducing fluctuations caused by localized states. The background inverse compressibility between integer quantum Hall states is close to zero at $4<\nu<8$, but is markedly more negative for $0<\nu<4$. The more negative inverse compressibility in the lowest LL is consistent with previous observations in monolayer graphene \cite{Geimcap,befSci,Skinner}.

Even within the lowest LL, the background inverse compressibility has two different characteristic shapes. It is more negative and less flat when starting to fill from an even filling factor than when starting to fill from an odd filling factor. One might expect that as states are filled from an even filling factor, electron-electron interactions break the degeneracy between the $N=0$ and $N=1$ LLs, so that only the $N=0$ LL is occupied. When states are filled beyond an odd filling factor, the $N=0$ LL is already full, so electrons start to occupy the $N=1$ LL. The more negative background inverse compressibility between $\nu=0$ and $1$ and between $\nu=2$ and $3$ points to the presence of less screening or more electron-electron correlations in the $N=0$ LL when compared with the $N=1$ LL with a underlying filled $N=0$ LL.

\begin{figure}
\centering
\includegraphics[width=\textwidth]{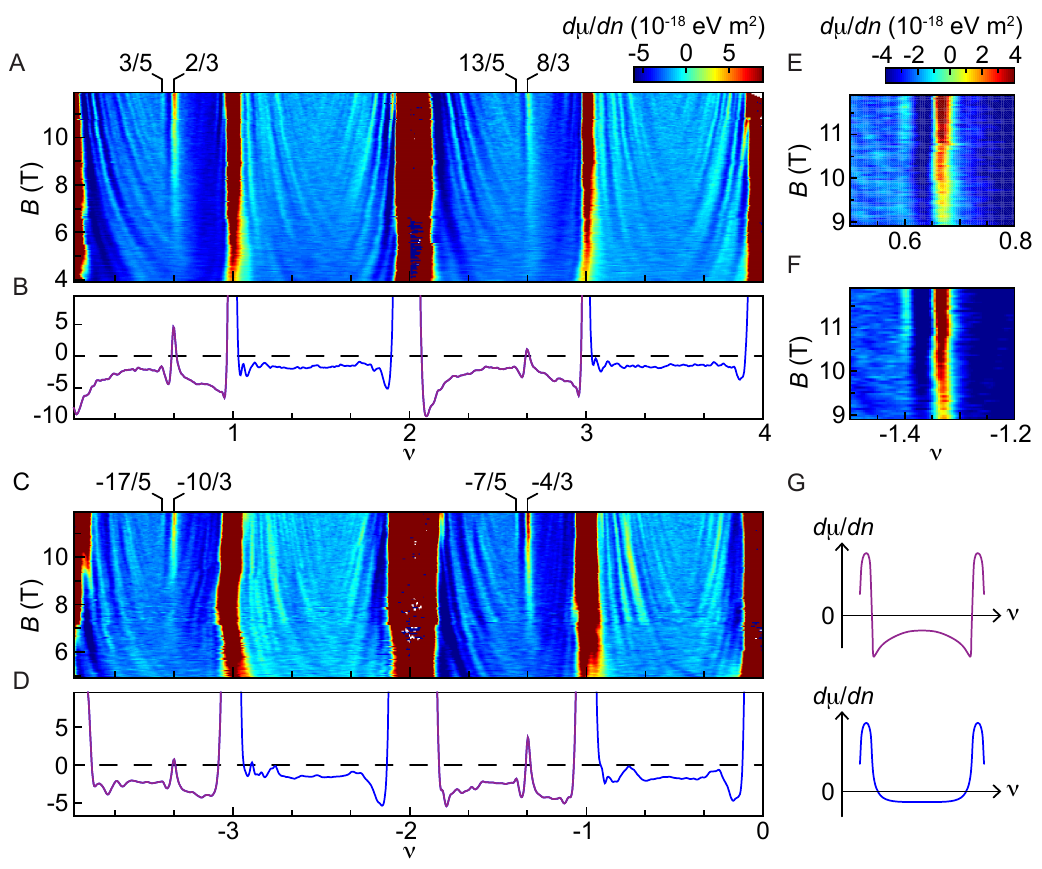}
\caption{
\label{BNGFig2}}
\end{figure}

Figures \ref{BNGFig2}A and \ref{BNGFig2}C show the inverse compressibility as a function of filling factor and magnetic field after we further cleaned the sample by current annealing. Quantum Hall states appear as vertical features when plotted in this form, while localized states, which occur at a constant density offset from their parent states, curve as the magnetic field is changed \cite{befSci, Ilani_Nature427_328_2004}. We can then unambiguously identify the incompressible states that appear at $\nu=-10/3, -4/3, 2/3,$ and 8/3 as FQH states. Interestingly, these states follow a $\nu=2p+2/3$ sequence. Above 10~T, we also see evidence of developing states at $\nu=-17/5, -7/5, 3/5$ and 13/5, which follow a similar $\nu=2p+3/5$ sequence. Zoom-ins of the regions around $\nu= 3/5$ and $-7/5$ are shown in Figs. \ref{BNGFig2}E and \ref{BNGFig2}F, respectively. The FQH states closer to the charge neutrality point are more incompressible than those at higher filling factors, and they persist to fairly low magnetic fields, with the last hints disappearing around 6~T.

The lineplots in Figs. \ref{BNGFig2}B and \ref{BNGFig2}D show the average inverse compressibility in each filling factor range from 7.9 to 11.9~T. They highlight the FQH states identified above as well as the different behaviors exhibited by the inverse compressibility, which shows especially strong divergences as filling factor is increased from $\nu=0$ and $\nu=2$. Similar to $\nu>0$, the background inverse compressibility at negative filling factors also displays the same even-odd effect with more negative values when starting to fill from an even filling factor. We note that the FQH states coincide with areas of more negative background inverse compressibility. This correlation is consistent with previous experiments where a negative background compressibility was attributed to Coulomb interactions between charge carriers \cite{Eisenstein, Geimcap}. Despite theoretical predictions of robust FQH states in the $N = 2$ LL and experimental hints in other samples \cite{Maher}, we do not observe any FQH states between $|\nu| = 4$ and 8.

Remarkably, the observed sequence of FQH states and the background inverse compressibility pattern break particle-hole symmetry and instead follow a $\nu \rightarrow \nu+2$ pattern. The $\nu \rightarrow \nu+2$ symmetry that we observe indicates that the orbital degeneracy uniquely present in bilayer graphene is playing an important role. Recent theoretical work on the FQH effect in the lowest LL has predicted the presence of FQH states with a $\nu \rightarrow \nu+2$ symmetry in bilayer graphene based on a model that incorporates the strong screening and LL mixing present in the lowest LL of bilayer graphene \cite{Papic_arXiv1307_2909}. The absence of FQH states between $\nu=-3$ and $-2$ as well as its $\nu \rightarrow \nu+2$ symmetric counterparts suggests a difference in electron-electron interactions between partial filling when both the $N=0$ and $N=1$ LLs are empty and partial filling of the $N=1$ LL when the $N=0$ LL is full. The increased LL mixing present when the $N=0$ LL is full \cite{Peterson} may be weakening the strength of FQH states in the $N=1$ LL.

Our observed FQH sequence also suggests possible orbital polarization of the FQH states. The FQH states we observe at $\nu=2p+2/3$ could be "singlet" states of $N=0$ and $N=1$ orbitals, or could arise from a 2/3 state with full orbital polarization. The next strongest FQH states we observe occur at $\nu=2p+3/5$, which must have some nonzero orbital polarization. It is worthwhile to note that the strongest FQH states at multiples of $\nu=1/5$ do not have even numerators, in contrast with recent theoretical predictions \cite{Papic_arXiv1307_2909}.

The FQH states that we observe are different from those seen in previous experiments on bilayer graphene, which may point to different patterns of symmetry-breaking in the different systems. Ki and coworkers observed FQH states at $\nu=-4/3,$ and $-1/2$, with hints of additional features at $\nu=-8/5$ and $-2/3$ \cite{Morpurgo}; the only FQH state seen in both devices is $\nu=-4/3$. It is also possible that the effective interactions present in the two samples may be different due to differences in screening between a suspended bilayer and a bilayer on a substrate. The fact that different sample preparations result in different FQH states could be a sign of the theoretically predicted tunability of the FQH effect in bilayer graphene \cite{PapicTun, Snizhko, Apalkov}. Future experiments in which a perpendicular electric field and/or a parallel magnetic field are applied to the sample may provide insight into the conditions under which different FQH states are favored. We can contrast this with monolayer graphene, where both suspended and substrate-supported samples have shown similar sequences of FQH states \cite{Du, Bolotin, Dean, Lee, befSci, Javier}. Though phase transitions of FQH states were observed \cite{befPhase} in monolayers, these involved only changes in the spin and/or valley polarization, and did not change the sequence of observed FQH states. In bilayer graphene, however, it appears possible to experimentally tune the relative strengths of various incompressible FQH states.

\begin{figure}
\centering
\includegraphics[width=\textwidth]{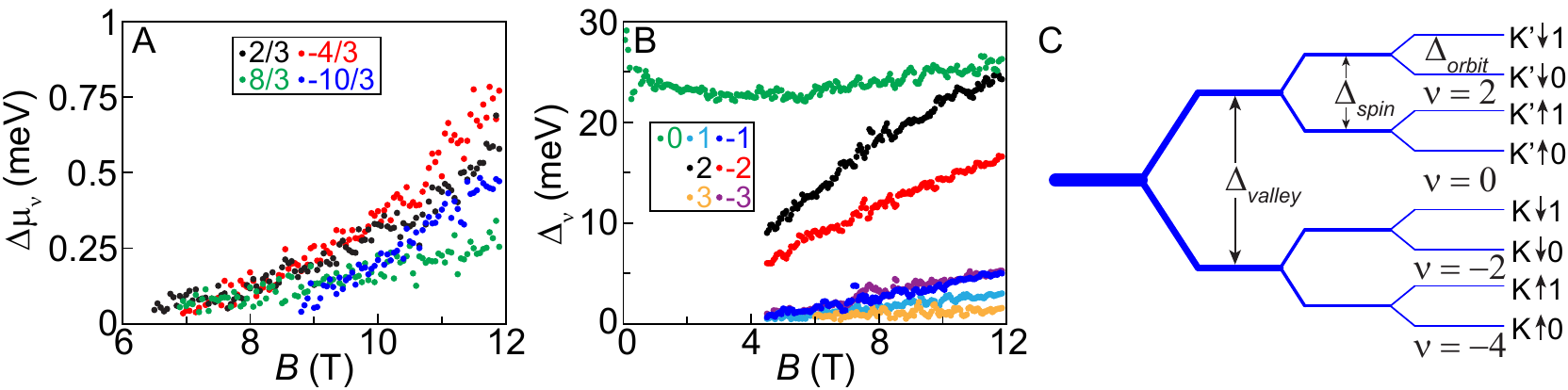}
\caption{
\label{BNGFig3}}
\end{figure}

We can integrate the inverse compressibility as a function of density to obtain the energy cost of adding an electron to the system, as discussed in ref. \cite{befSci}. This quantity must be divided by the quasiparticle charge associated with each state to determine the corresponding energy gap $\Delta_{\nu}$. The most likely quasiparticle charge for states at multiples of $\nu=1/3$ is $e/3$, but the nature of the FQH states in bilayer graphene is not yet fully understood, so we plot the extracted steps in chemical potential $\Delta\mu_{\nu}$ in Fig. \ref{BNGFig3}A. For $\nu=-4/3$ and $\nu=2/3$, $\Delta\mu_{\nu}$ is about 0.75 and 0.6~meV, respectively, at $B=12$~T. Assuming a quasiparticle charge of $e/3$, the energy gap we find at $\nu=-4/3$ is comparable with, if somewhat larger than, that found in ref. \cite{Morpurgo} at similar magnetic fields. The gaps of FQH states further away from charge neutrality are smaller; $\Delta\mu_{-10/3}$ and $\Delta\mu_{8/3}$ are only about 0.5 and 0.3~meV at $B=12$~T. All of the extracted gaps appear to scale approximately linearly with $B$, but a $\sqrt{B}$-dependence may also fit the gaps at $\nu=8/3$ and $\nu=-10/3$. Previous measurements of broken-symmetry integer states in suspended bilayers found a linear-$B$ dependence of the gaps, which was attributed to LL mixing \cite{Jens, Nandkishor_PRL104_156803_2010}.

The energy gaps of the integer filling factors $|\nu|<4$ are shown in Fig. \ref{BNGFig3}B. All of the gaps increase with $B$, except for $\nu=0$, which is fairly constant around 23-25~meV over almost the full range in magnetic field. Around 4~T, the gap dips slightly before increasing again at $B=0$~T. The size of the gap and its persistence to zero field lead us to conclude that the ground state at $\nu=0$ is layer-polarized. If we assume that the $\nu \rightarrow \nu+2$ symmetry arises from the orbital degree of freedom, we can fully determine the sequence of symmetry-breaking in the sample: valley polarization is first maximized, then spin polarization, and finally orbital polarization, as illustrated in Fig. \ref{BNGFig3}C. The large valley polarization in our sample relative to other bilayer devices may be caused by interactions with the substrate. Large band gaps have been observed in monolayer graphene samples on h-BN with a proximal gate, which have been attributed to the breaking of sublattice symmetry by h-BN or screening from the nearby metal \cite{Javier, DGG,Ponomarenko}. It is also possible that the difference in distance between the top layer to the graphite gate and the bottom layer to the graphite gate is creating a potential difference in the two layers \cite{McCannGap}, or that the different environments experienced by each layer play a role. Even if the $\nu=0$ gap is caused by single-particle effects, its constancy over our entire field range is somewhat surprising because both the potential difference between the layers and the Coulomb energy are expected to contribute to the gap \cite{Toke}.

\begin{figure}
\centering
\includegraphics[width=\textwidth]{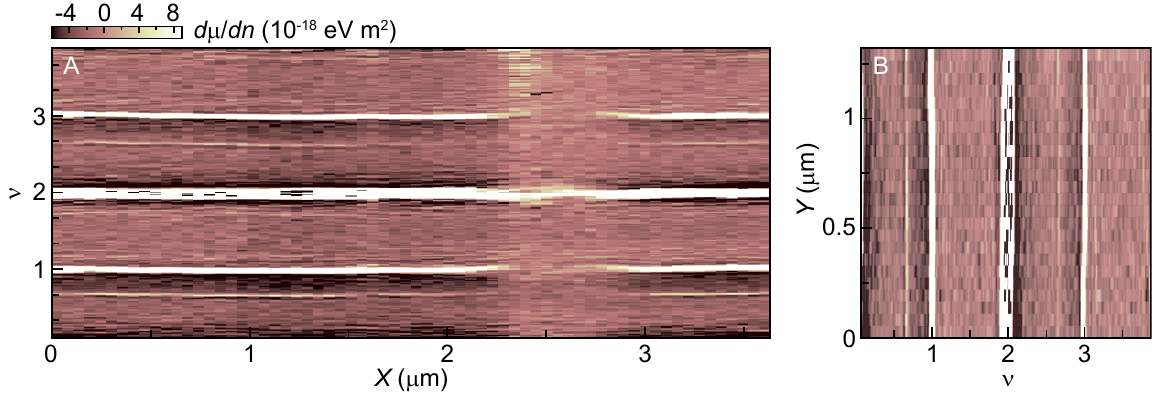}
\caption{
\label{BNGFig4}}
\end{figure}

All of the measurements described above were performed at a single location on the sample. Local measurement allows us to find the cleanest regions and study the properties of FQH states in those areas. In addition, the local nature of our probe allows us to, in effect, measure multiple independent samples by measuring at different locations. Figure \ref{BNGFig4} shows linescans of inverse compressibility as a function of filling factor and position. The net level of doping remains fairly constant over the entire range of these spatial scans, but there are fluctuations in the strengths of the broken-symmetry and FQH states, likely due to differences in the amount of local disorder. Despite these fluctuations, the overarching pattern of FQH states is consistent across the entire sample, and also did not change with current annealing. The electron-hole asymmetric sequence of FQH states can therefore be attributed to the intrinsic properties of bilayer graphene, rather than disorder. The observation of an unconventional sequence of FQH states in bilayer graphene indicates the importance of its underlying symmetries and opens new avenues for exploring the nature and tunability of the FQH effect.\\

\noindent
\textbf{References and Notes:}
{\def\section*#1{}

}

\noindent
\textbf{Acknowledgments:} We would like to thank D. A. Abanin, A. H. MacDonald, B. Skinner, and I. Sodemann for useful discussions. We also thank J. D. Sanchez-Yamagishi for help with fabrication. This work is partially supported by the US Department of Energy, Office of Basic Energy Sciences, Division of Materials Sciences and Engineering under Award \#DE-SC0001819. Device optimization and initial measurements were supported by the 2009 U.S. ONR Multi University Research Initiative (MURI) on Graphene Advanced Terahertz Engineering (Gate) at MIT, Harvard, Boston University, and the NRI-SWAN program. This material is based in part upon work supported by the National Science Foundation under Cooperative Agreement No. DMR-1231319. Devices were fabricated at the Center for Nanoscale Systems at Harvard University, supported by NSF under grant ECS-0335765. A.K. acknowledges support from the Department of Energy Office of Science Graduate Fellowship Program (DOE SCGF). A.J.L. is supported through a fellowship from the Department of Defense (NDSEG program).

\clearpage

\noindent
\textbf{Fig. 1}. Sample image and characterization. (\textbf{A}) Optical image of the device with colored overlays showing the graphite (blue), boron nitride (purple), and monolayer-bilayer graphene hybrid (red). The black dashed line marks the interface between monolayer and bilayer. The scale bar is 2~$\mu$m. (\textbf{B}) Inverse compressibility $d\mu/dn$ as a function of filling factor $\nu$ at magnetic field $B = 2$~T. (\textbf{C}) Average inverse compressibility between $B = 8$ and 11.5~T as a function of filling factor after current annealing to 4V. Shaded areas indicate regions of more negative background inverse compressibility.

\clearpage

\clearpage

\noindent
\textbf{Fig. 2}. Fractional quantum Hall states in bilayer graphene. (\textbf{A}) and (\textbf{C}) Inverse compressibility as a function of filling factor and magnetic field. The color scales are the same in both panels. (\textbf{B}) and (\textbf{D}) Average inverse compressibility between $B = 7.9$ and 11.9~T as a function of filling factor. Colors indicate regions of similar behavior in the background inverse compressibility. (\textbf{E}) and (\textbf{F}) Inverse compressibility as a function of filling factor and magnetic field near $\nu=-7/5$ and 3/5. (\textbf{G}) Schematic diagram highlighting the differences in background inverse compressibility between $\nu=2p$ and $\nu=2p+1$ in purple and $\nu=2p+1$ and $\nu=2p$ in blue.\\

\clearpage

\noindent
\textbf{Fig. 3}. Steps in chemical potential. (\textbf{A}) Steps in chemical potential of the fractional quantum Hall states as a function of magnetic field. (\textbf{B}) Energy gaps of the integer broken-symmetry states in the lowest Landau level. (\textbf{C}) Schematic diagram showing the order of symmetry breaking in the sample.

\clearpage

\noindent
\textbf{Fig. 4}. Spatial dependence of fractional quantum Hall features. Inverse compressibility as a function of filling factor and position $X$ (\textbf{A}) and $Y$ (\textbf{B}). The sequence of FQH states does not vary with position.

\clearpage

\noindent
\textbf{Supplementary Materials:}
\setcounter{figure}{0}

\noindent
Materials and Methods

\noindent
Figures S1-S2\\

\noindent
\textbf{Supplementary Materials:}

\noindent
\textbf{Effects of current annealing}
\indent

The data presented in the main text were taken after multiple rounds of current annealing, with a maximum source-drain voltage $V_{sd} = 10$~V. Figure \ref{BNGFig6} shows the progression of the behavior before and after current annealing. Prior to current annealing the sample, we observed all the integer broken-symmetry states between $\nu = \pm 4$, but no FQH states were apparent. After current annealing to 4~V, we observed hints of incompressible states at $\nu=2/3$ and $\nu=8/3$ [Fig \ref{BNGFig6}A]. These states became more robust after annealing to 8~V [Figs. \ref{BNGFig6}B-D] and additional FQH states appeared at $\nu = -4/3$ and $-10/3$. Throughout all current annealing steps, the sequence of incompressible FQH states did not change; current annealing appears to only increase their strength. It is worthwhile to note that the steps in chemical potential were slightly larger for $\nu = 2/3$ and 8/3 after annealing to only 8~V than the data presented in Fig. \ref{BNGFig3}A.
\\

\noindent
\textbf{Conversion of gate voltage to filling factor with large geometric capacitance}
\indent

The proximity of the graphite gate resulted in a large geometric capacitance, $C_G$, to the sample, causing the total capacitance, $C_T$, to depend strongly on the compressibility of the sample. Incompressible states dramatically alter $C_T$ because quantum capacitance dominates over geometric capacitance in this regime. Figure \ref{BNGFig7} shows the compressibility as a function of back-gate voltage, $V_{bg}$, and the very wide integer quantum Hall states clearly demonstrate the filling-factor dependent change in the total capacitance.

In order to assign filling factor, we use the distance between $\nu=2/3$ and 1 to find $C_T$ in the $\nu=0$ to 1 range. The equation,
\begin{equation}
\frac{1}{C_T}=\frac{1}{C_G} + \frac{1}{e^2}\frac{d\mu}{dn},
\end{equation}
where $d\mu/dn$ is the average background inverse compressibility between $\nu=0$ and 1, allows us to determine $C_G$. The extracted $C_G$ corresponds to a back-gate to density conversion ratio of 2.7x10$^{16}$~cm$^{-2}$/V, which is reasonable given a h-BN thickness of 7.5-8~nm and a dielectric constant a bit less than 4.

To find $C_T$ in a different filling factor range, we then use Equation 1, the average compressibility in this new filling factor range, and $C_G$, which we assume is constant. The extracted $C_T$ then provides the conversion between the back-gate voltage and density. Throughout the manuscript, we use the same $C_T$ for $-4<\nu<-3, 0<\nu<1$, and their $\nu \rightarrow \nu+2$ analogues. We also use a single $C_T$ for filling factors between $\nu=-3$ and $-2$, $\nu=1$ and 2, and their $\nu \rightarrow \nu+2$ counterparts. For $|\nu|>4$, we use the geometric capacitance.
\\

\noindent
\textbf{Materials and Methods}

\indent
Graphite was mechanically exfoliated onto an O$_2$ plasma-cleaned doped Si wafer capped with 285~nm of SiO$_2$. Suitable graphite pieces were found using a combination of optical microscopy and atomic force microscopy (AFM). A 7.5~nm thick piece of hexagonal-boron nitride (h-BN) was then transferred onto the graphite using the process detailed in ref. \cite{Javier}. A hybrid monolayer-bilayer graphene flake was then transferred onto the h-BN using the same method. Contacts were defined to the graphene and graphite using electron-beam lithography before thermal evaporation of Cr/Au (1~nm/85~nm) contacts and liftoff in warm acetone. The sample was cleaned in a mixture of Ar/H$_2$ at 350 $^{\circ}$C before each transfer step and after liftoff, and it was further cleaned using an AFM tip. Measurements were made in a $^3$He cryostat at approximately 450~mK. The sample was cleaned in the cryostat using current annealing. The sample measures 8 $\mu$m from contact to contact and is 4 $\mu$m wide. All measurements presented here were made on the bilayer side of the flake.

To fabricate the scanning SET tip, a fiber puller was used to make a conical quartz tip. Al leads (16 nm) were evaporated onto either side of the quartz rod, and following an oxidation step, 7 nm of additional Al was evaporated onto the tip to create the island of the SET. The diameter of the SET is approximately 100 nm, and it was held 50-100 nm above the graphene flake during measurements. Compressibility measurements were performed using AC and DC techniques similar to those described in refs. \cite{befSci, Jens}. The SET serves as a sensitive measure of the change in electrostatic potential $\delta \phi$, which is related to the chemical potential of the graphene flake by $\delta \mu = -e \delta \phi$ when the system is in equilibrium. In the AC scheme used to measure $d\mu/dn$, an AC voltage is applied to the graphite gate to weakly modulate the carrier density of the flake, and the corresponding changes in SET current are converted to chemical potential by normalizing the signal with that of a small AC bias applied directly to the sample. For DC measurements, a feedback system was used to maintain the SET current at a fixed value by changing the tip-sample bias. The corresponding change in sample voltage provides a direct measure of $\mu(n)$.

\clearpage

\renewcommand*{\thefigure}{S\arabic{figure}}
\begin{figure}
\centering
\includegraphics[width=\textwidth]{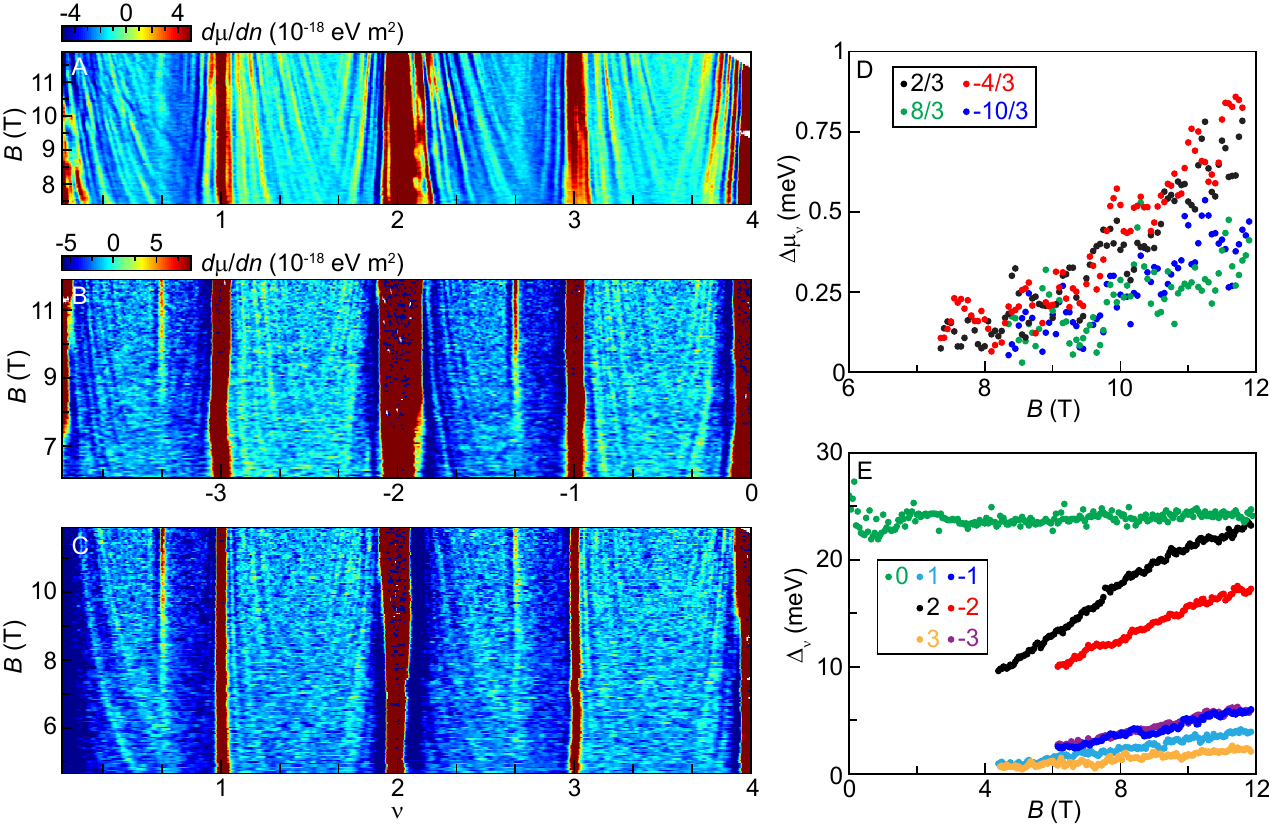}
\caption{
\label{BNGFig6}}
\end{figure}

\noindent
\textbf{Figure S1}. Effects of current annealing. (\textbf{A}) Inverse compressibility as a function of filling factor and magnetic field after annealing to a source-drain voltage $V_{sd} = 4$~V. Weak incompressible peaks are visible at $\nu = 2/3$ and 8/3. (\textbf{B}) and (\textbf{C}) Inverse compressibility as a function of filling factor and magnetic field after annealing to $V_{sd} = 8$~V. The FQH states become sharper and more incompressible, and they persist to lower magnetic field. (\textbf{D}) Steps in chemical potential of the FQH states after annealing to 8~V. (\textbf{E}) Energy gaps of the integer quantum Hall states after annealing to 8~V.\\

\clearpage

\begin{figure}
\centering
\includegraphics[width=\textwidth]{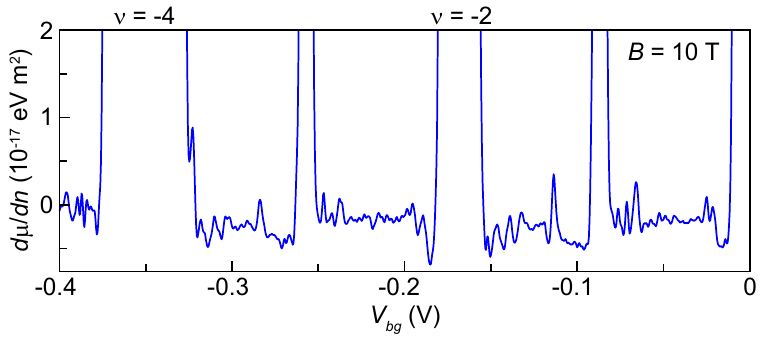}
\caption{
\label{BNGFig7}}
\end{figure}

\noindent
\textbf{Figure S2}. Effect of large geometric capacitance. Inverse compressibility as a function of back gate voltage $V_{bg}$. For incompressible states, the quantum capacitance dominates over geometric capacitance, leading to very broad incompressible peaks, particularly at $\nu = -4$ and $-2$. The spacing between integer quantum Hall states also varies due to differences in background compressibility.

\end{document}